\def\putfigure{0}	
\documentstyle[12pt,aasms4]{article}
\def\gtwid{\mathrel{\raise.3ex\hbox{$>$\kern-.75em\lower1ex\hbox{$\sim$}}}}
\def\ltwid{\mathrel{\raise.3ex\hbox{$<$\kern-.75em\lower1ex\hbox{$\sim$}}}}
\def\\{\hfil\break}
\def\ie{{\it i.e.\ }}
\def\eg{{\it e.g.\ }}
\def\etal{{\it et al.\ }}
\def\hmpc{$h^{-1}$Mpc}

\begin{document}
\title{A New Approach to Probing Large Scale Power with Peculiar Velocities}
\vskip 0.5cm
\author{\bf Hume A. Feldman\footnote{feldman@ukans.edu}}
\vskip 0.3cm
\affil{{\it Department of Physics \& Astronomy, University of Kansas\\
Lawrence, KS 66045}}
\vskip 0.3cm
\author{\bf Richard Watkins\footnote{watkins@mojo.dartmouth.edu}}
\affil{\it Department of Physics \& Astronomy, Dartmouth College\\ Hanover, NH 03755}

\baselineskip 12pt plus 2pt

\begin{abstract}
We propose a new strategy to probe the power spectrum on large scales
using galaxy peculiar velocities.  We explore the properties of surveys
that cover only two small fields in opposing directions on the sky.
Surveys of this type have several advantages over those that attempt to
cover the entire sky; in particular, by concentrating galaxies in narrow
cones these surveys are able to achieve the density needed to measure
several moments of the velocity field with only a modest number of
objects, even for surveys designed to probe scales $\gtwid 100$\hmpc.
We construct mock surveys with this geometry and analyze them in terms
of the three moments to which they are most sensitive.  We calculate
window functions for these moments and construct a $\chi^2$ statistic
which can be used to put constraints on the power spectrum.  In order to
explore the sensitivity of these surveys, we calculate the expectation
values of the moments and their associated measurement noise as a
function of the survey parameters such as density and depth and for
several popular models of structure formation.  In addition, we have
studied how well these kind of surveys can distinguish between different
power spectra and found that, for the same number of objects, cone
surveys are as good or better than full-sky surveys in distinguishing
between popular cosmological models.  We find that a survey with
$200-300$ galaxy peculiar velocities with distance errors of $15\%$ in
two cones with opening angle of $\sim 10^\circ$ could put significant
constraints on the power spectrum on scales of $100-300$\hmpc, where few
other constraints exist.  We believe that surveys of the kind we
describe her could provide a valuable tool for the study of large scale
structure on these scales and present a viable alternative to full sky
surveys.
\end{abstract}

\noindent{\it Subject headings}: cosmology: distance scales -- cosmology: large
scale structure of the universe -- cosmology: observation -- cosmology:
theory -- galaxies: kinematics and dynamics -- galaxies: statistics

\def\figureone{In the top three panels we plot the window
functions associated with the moments $U_1$, $U_2$, and $U_3$ for
different values of $R_o$, the location of the peak of the radial
selection function for mock surveys of 200 galaxies.  In the bottom
panel we see the power spectra we used in our analysis. To find the
expectation values of the different moments we integrate the product of
the window functions of the top three panels with each of the power
spectra.}
\def\figuretwo{Here we see the values of $Q_i$ 
for different power spectra as a function of the number of objects in
the survey. $Q_i$ rises with the number of points as expected as the
noise inherent in the measurement becomes smaller with increased
density.}
\def\figurethree{The $\chi^2$ for 3 degrees of freedom for 
HCDM normalized to COBE given a Universe with $\Lambda$CDM, and vice
versa, for $10^{\circ}$ cone surveys, with the three lowest moments and
for a full-sky survey (with the zone-of-avoidance removed) using the
three bulk flow components as the three degrees of freedom, as a
function of the number of objects.  The full--sky surveys have the same
radial distribution and errors as the cone surveys.  We also show the
$\chi^2$ for COBE normalized CDM and cluster normalized CDM.  From the
figure we see that the three moments calculated for the cone survey do
better than the three bulk-flow components of the full-sky survey at
distinguishing between the models for the same number of observed
galaxies.}

\section{Introduction}
\baselineskip 23pt plus 2pt

One of the most important goals of cosmology is to determine the power
spectrum of initial density fluctuations in the Universe.  The primary
tools in this endeavor have been redshift surveys, which have been used
to probe scales $\ltwid 100$ \hmpc, and microwave background anisotropy
measurements, which tell us about the power spectrum on scales $\gtwid
1000$ \hmpc\ (we use $h=H_o/(100$ km/s/Mpc) where $H_o$ is the Hubble
constant). The gap between these scales, where we have relatively little
knowledge of the power spectrum, is notable in several respects.  First,
theoretical considerations tell us that the power spectrum should have a
maximum in this region at a scale corresponding to the horizon size at
the time that the Universe became matter dominated.  While redshift
surveys have hinted that the power spectrum does indeed turn over at a
scale $\sim200$ \hmpc\ (\eg Fisher \etal 1993, Feldman \etal 1994), the
errors inherent in the measurement are too large to be definitive.
Further, several recent studies have suggested that there might be much
more power on these scales than is usually assumed in models of large
scale structure formation (Broadhurst \etal\ 1990, Landy \etal\ 1996,
Ainasto \etal\ 1996.)  If these suggestions are correct it will have
important ramifications for our understanding of how structure formed in
the Universe.

There are several observations planned or in progress that will attempt
to study the power spectrum in this regime.  From above, there are the
small angle microwave anisotropy measurements. From below,
the next generation redshift observations should
significantly extend the range over which surveys can reliably
determine the power spectrum.  However, it may be a decade or longer
before these ambitious and complex projects will produce measurements of
the power spectrum on scales $\sim 200$ \hmpc.

A third method of measuring the power spectrum on large scales involves
the study of large scale motions of galaxies.  There were several recent
attempts to determine the power spectrum from observations of peculiar
velocities.  Kolatt \& Dekel (1997) use the POTENT reconstruction of the
density field to determine P(k). Zaroubi \etal\ (1996) estimated the
power spectrum from the Mark III catalog of peculiar velocities using
Bayesian statistics.  One advantage of this method is that the large
scale velocity field probes the matter distribution in the Universe
directly, and not merely the light distribution as redshift surveys do.
However, the errors in velocity estimates are typically some fraction of
the distance of the sample points, which in the case of distant objects
can mean that the errors are larger than the peculiar velocity being
measured.  This is partially rectified by measuring only the lowest
moment of the velocity field, namely the bulk flow.  Since the bulk flow
is, in a sense, an average of the velocities in the sample, its error is
reduced over that of an individual measurement by the square root of the
number of objects.

Two recent efforts to measure the bulk velocities of large volumes have
resulted in contradictory conclusions. 
Lauer \& Postman (LP 1994) found that the inertial
frame defined by Abell clusters within $15,000$km/s exhibits a bulk
velocity of $\approx 700$ km/s with respect to the cosmic microwave
background (CMB) rest frame.  More recently, Riess \etal (RPK 1995) used
type Ia supernovae to measure the bulk velocity of
a similar volume; their results are consistent with their sample being
at rest relative to the CMB rest frame.  In recent analyses (Feldman \&
Watkins 1994, Strauss \etal 1995, Jaffe \& Kaiser 1995, Watkins \&
Feldman 1995) it was shown that both power spectra from structure
formation models as well as results from redshift surveys are
inconsistent with the LP measurement at the $2-3\sigma$ level, whereas
they are quite consistent with the RPK result.  Furthermore, the RPK and
LP results seem to be inconsistent with each other at a high confidence
level.  

Abell clusters and type Ia supernovae are rare, thus the
LP and RPK samples are by necessity quite sparse.  This
limits their ability to accurately measure moments of the velocity field
beyond the bulk flow.  In addition,
obscuration by our Galaxy strongly
constrains how accurately these surveys can ascertain components of the
bulk flow in the plane of the Galaxy.  Thus the only component of the
bulk flow that LP and RPK are able to report with a reasonable
significance is that along the Galactic poles.  In this {\it
Letter} we propose an alternative approach to gathering full sky surveys
of peculiar velocities; we explore the possibility of probing the power
spectrum on large scales by measuring the velocities of galaxies in only
small patches of the sky.  While this survey will necessarily be 
sensitive to only one component of the bulk flow, the increase in the {\it
density} of objects will allow more accurate measurement of higher
moments of the velocity field.  
An added advantage of this approach is that the higher moments
probe a different range of scales than the bulk flow, making it possible
to constrain the power spectrum more precisely.

We examine several factors necessary for the design of a useful peculiar
velocity survey.  We
concentrate on surveys which cover two fields directly opposite to each
other.  These surveys can probe larger scales and
are less susceptible to radial biases than one-sided surveys.  We
construct mock catalogs for these surveys and show how their sensitivity
depends on depth and the number of objects measured.  In particular, we
calculate the expectation values and expected errors associated with the
three most easily measured moments assuming several different power
spectra.

\section{Analysis}

As stated above, individual velocity measurements are too noisy to allow
us to accurately map the peculiar velocity field $\vec v(\vec r)$.
Instead we expand $\vec v(\vec r)$ in a given region in terms of its
moments (Kaiser 1991, Jaffe \& Kaiser 1995),
\begin{equation} 
v_i(\vec r) = u_i + r_j p_{ij} + r_j r_k q_{ijk}+\dots
\end{equation} 
Here, $u_i$ and $p_{ij}$ are usually referred to as the bulk flow and
the shear tensor respectively.

In practice we can measure only the radial component of velocities.
Thus if our objects lie in a cone around a direction given by $\hat R$,
we will be most sensitive to the component of the velocity along $\hat
R$, $v_R= \vec v\cdot \hat R$.  For this situation it is sufficient to
model the velocity as being entirely along the $\hat R$ direction and
depending only on $\vec r\cdot \hat R$, giving us a much simpler
expansion,
\begin{equation}
v_R(\vec r\cdot \hat R)= U_1 + U_2(\vec r\cdot\hat R/R_o) + U_3(\vec
r\cdot\hat R/R_o)^2,
\end{equation}
where we have introduced the arbitrary scale $R_o$ in order that the
constants $U_i$ will all have the same units.  In most of the analysis
below we choose $R_o=100$\hmpc.  The constants $U_i$ represent the one
component each of $u_i$, $p_{ij}$, and $q_{ijk}$ to which these surveys
are most sensitive.

In terms of this model, the estimated line--of--sight velocity $S_n$
measured for the $n$th galaxy at a position $\vec r_n$ can be written as
$S_n = \sum_{i=1}^3 F_{n,i}\ U_i + \epsilon_n$, where $F_{n,i}= (\hat
r_n\cdot \hat R)(\vec r_n\cdot \hat R/ R_o)^{i-1}$.  Here we assume that
the noise $\epsilon_n$ is drawn from a Gaussian with zero mean and
variance $\sigma_n^2 + \sigma_*^2$, where $\sigma_n$ is the estimated
uncertainty in the measurement of the line-of-sight velocity and
$\sigma_*$ is introduced to account for contributions to the velocity of
the galaxies in the survey arising from nonlinear effects as well as
from the components of the velocity field that we have neglected in our
model (Kaiser 1988).  We shall take $\sigma_* =400$ km/s for all of our
calculations, although in practice this choice makes little difference
given the large values of $\sigma_n$ for most of the galaxies that we
will consider.

Given a sample of objects with positions $\vec r_n$ and line-of-sight
velocities $S_n$, the maximum likelihood solutions for the constants
$U_i$ are given by
\begin{equation}
U_i = (A^{-1})_{ij}\sum_n {F_{n,j}\ S_n\over (\sigma_n^2 +
\sigma_*^2)},
\end{equation}
where the matrix $A_{ij} = \sum_{n} {F_{n,i}\ F_{n,j}\over (\sigma_n^2 +
\sigma_*^2)}$; here and below we use the Einstein summation convention
for repeated indices.

The theoretical expectations for the constants $U_i$ can be expressed in
the form of a covariance matrix,
$
R_{ij}= \langle U_iU_j\rangle=R^{(v)}_{ij} + (A^{-1})_{ij}\ 
$
where $R^{(v)}_{ij}$ is the contribution from the velocity field and
$(A^{-1})_{ij}$ is the noise.  The matrix $R^{(v)}_{ij}$ is obtained by
convolving the power spectrum with a window function,
\begin{equation}
R^{(v)}_{ij} = {1\over (2\pi)^3}\int P_{(v)}(k)W^2_{ij}(k)\ d^3k =
{H^2f^2(\Omega_o)\over 2\pi^2}\int P(k)W^2_{ij}(k)\ dk,
\end{equation}
where we have used the fact that in linear theory there is a simple
relationship between the velocity and density power spectrum,
$P_{(v)}(k)= (H^2/k^2)f^2(\Omega_o)P(k)$, with $f(\Omega_o)\approx
\Omega^{0.6}$.

The tensor window function $W^2_{ij}(k)$ is calculated from the
positions and velocity errors of the objects in the survey,
\begin{equation}
W^2_{ij}(k) = {1\over 4\pi}\int d\theta\ d\phi\ {\rm sin}\theta\
W_i(k)W^*_j(k),
\end{equation}
where
\begin{equation}
W_i(k) = (A^{-1})_{ij}\sum_{n} { {\hat k}\cdot{\hat r_n}\ F_{n,i}\
e^{i\vec k\cdot{\vec r_n}}
\over (\sigma_n^2 + \sigma_*^2)}.
\end{equation}

Once we have calculated $R_{ij}$ for a given power spectrum, we can
construct a $\chi^2$ statistic $\chi^2 = U_i\ R^{-1}_{ij}\ U_j$ for the
three degrees of freedom of the measured moments.
This statistic can be used to assess the compatibility of the power
spectrum with the measured $U_i$.  We note that the this statistic
properly accounts for the correlations between the moments, which can be
important for a sparsely sampled survey.  Alternatively, given a
parameterization of the power spectrum, $R_{ij}$ can be used in a
likelihood analysis to determine the most likely values of the
parameters (see, \eg, Jaffe \& Kaiser 1995).

In designing a survey, it is important to determine how well it can
distinguish between different power spectra.  We can quantify this by
calculating the expectation value of the $\chi^2$ for power spectrum
$P_A(k)$ given that the true power spectrum of the Universe is power
spectrum $P_B(k)$:
\begin{equation}
\langle\chi^2_{AB}\rangle = \langle (U_B)_i (R_A^{-1})_{ij} (U_B)_j\rangle
= (R_A^{-1})_{ij}\langle (U_B)_i (U_B)_j\rangle =
(R_A^{-1})_{ij}(R_B)_{ij}= Tr(R_A^{-1}R_B),
\end{equation}
where subscripts denote quantities calculated assuming power spectra
$P_A$ and $P_B$.  The value of $\langle\chi^2_{AB}\rangle$ can be used
to estimate the confidence level at which one
can rule out power spectrum $P_A$ in a universe with power spectrum $P_B$
given a particular catalog of objects and assumed velocity errors.

We constructed mock catalogs of fixed number of galaxies restricted to
two cones on opposite sides of the sky.  We found that the results were
roughly independent of the opening angle $\alpha$ of the cone as long as
$\alpha < 30^\circ$, giving some flexibility in the design of a survey.
The results we report in this {\it Letter} are for $\alpha = 10^\circ$.
Galaxies were distributed radially using a selection function obtained
by fitting an analytical function (cf. Feldman \etal 1994) to the radial
distributions of various magnitude limited Tully-Fisher surveys (we used
different surveys from the Mark III Catalog of Peculiar Velocities
[Willick \etal, 1995, 1996]).  We adjusted the depth of the survey by varying
the location of the peak of the distribution function, $R_o$, which is
directly related to the magnitude limit.  For definiteness, we assumed
the error in the measured velocity of a galaxy to be $15\%$ of its
distance, an error commonly reported for Tully-Fisher distance
measurements.  

Given a mock survey, we can calculate the window functions for the three
moments and determine which scales they are sensitive to.  Assuming a
power spectrum, we can also calculate the expected values of $U_i$ and
their associated noise. It is useful to define the parameters
\begin{equation}
Q_{i} \equiv {R_{ii}^{(v)}\over (A^{-1})_{ii}},
\label{eq:sn}
\end{equation}
that indicate how accurately a given survey can measure each of
the three moments.  Ideally we would like $Q_i\gtwid1$ for each
of the three moments.  The geometry and density of the survey
can be optimized to meet this requirement with the minimal
observational effort.  Although
$U_i$ can in principle be correlated,  we found that these correlations
are small.

\ifnum\putfigure<1
	\begin{figure}[hbt] \plotone{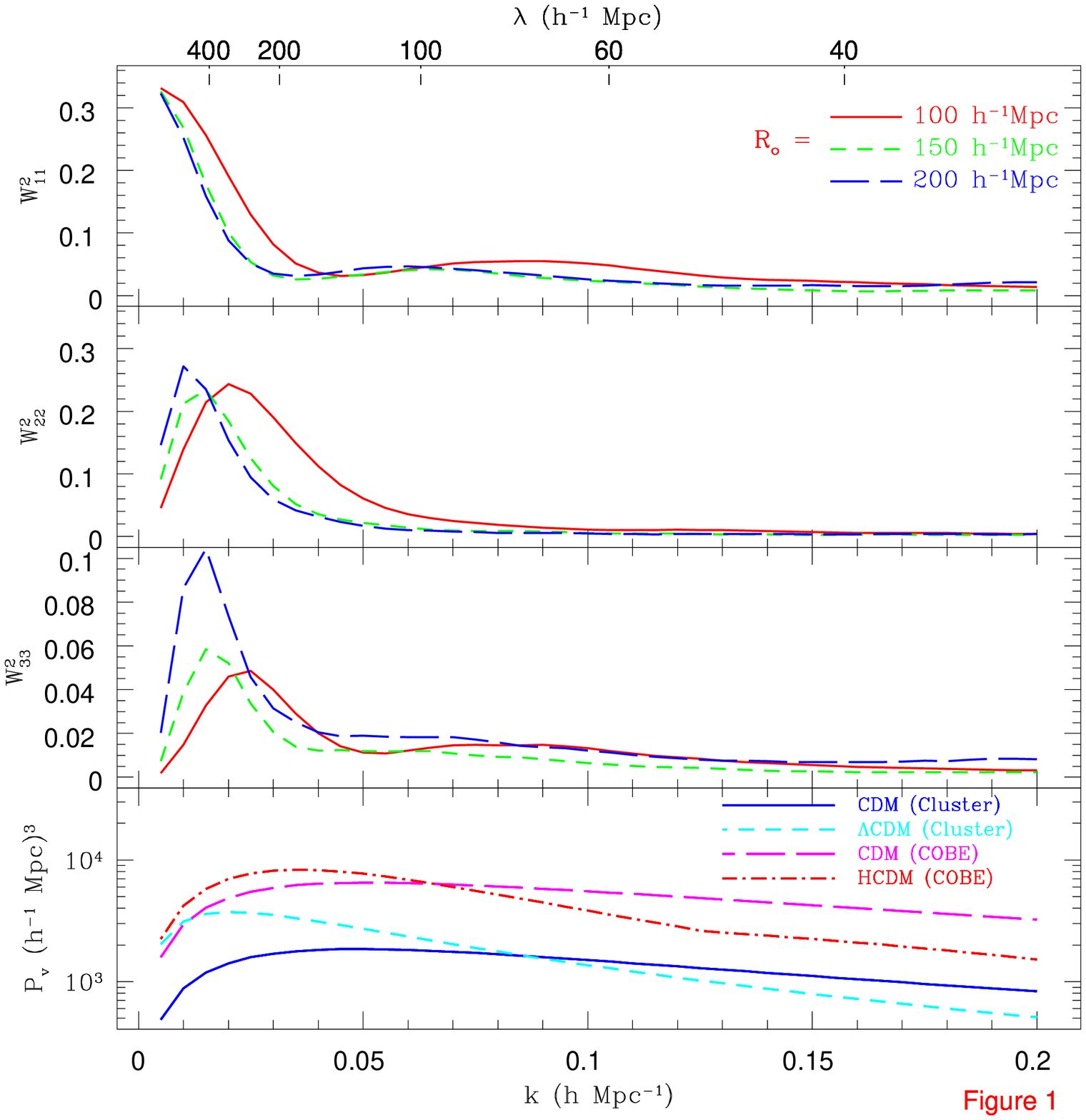} \caption{ \figureone }
	\label{fig1} \end{figure}
\fi

In Figure 1 we see a series of normalized window functions from three
mock surveys with different values of $R_o$, each with the same number
of galaxies ($n=200$).  While the window function for the bulk flow
($U_1$) has a maximum at $k=0$, the window functions for $U_2$ and $U_3$
are peaked at smaller scales.  Thus we see that the higher moments probe
a different region of the power spectrum than the bulk flow.  
Since we are looking only at a single component of the velocity
along the direction of the survey, the angular distribution of the
objects is relatively unimportant and edge effects are negligible.

In the bottom panel of Figure 1 we show the power spectra we used to
calculate expectation values for the moments; standard COBE normalized
CDM (Bardeen \etal\ 1986) and HCDM (Klypin \etal\ 1993) as well an
$\Omega=1$ CDM normalized the observed
abundance of clusters (see, \eg Eke \etal\ 1996)
and $\Lambda$CDM normalized to both cluster abundance and COBE.

\ifnum\putfigure<1
	\begin{figure}[hbt] \plotone{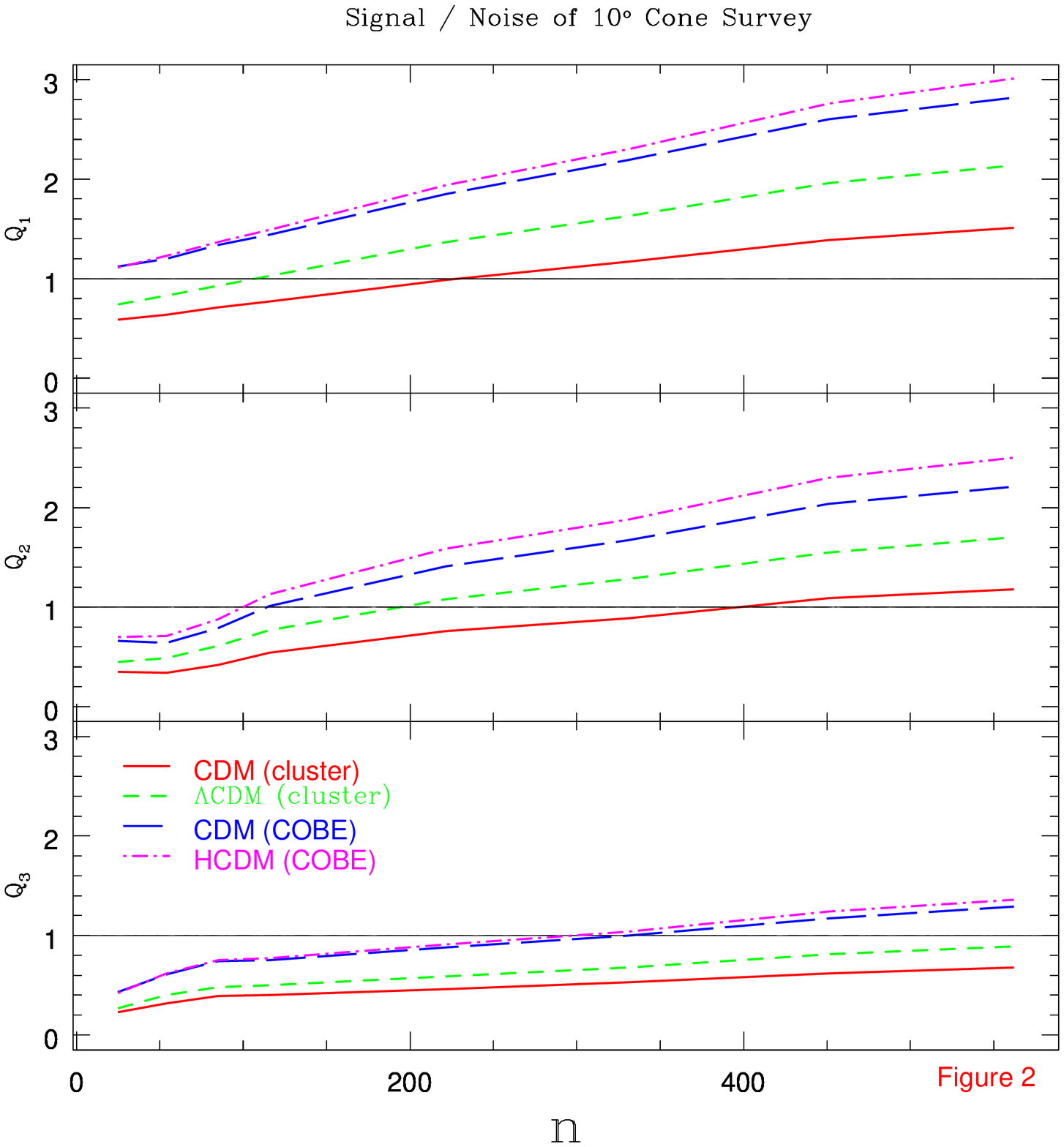} \caption{\figuretwo}
	\label{fig2} \end{figure}
\fi

In Figure 2 we show the $Q_i$ parameters (Eq. \ref{eq:sn}) for the three
moments for a survey with $R_o=100$\hmpc\ (effective depth
$\approx200$\hmpc) using the four power spectra described above.  We see
that we need some $200-300$ galaxies in our survey to get $Q_i\sim 1$
for the power spectra we consider that produce the largest velocities,
\ie COBE normalized CDM and HCDM.

\ifnum\putfigure<1
	\begin{figure}[hbt] \plotone{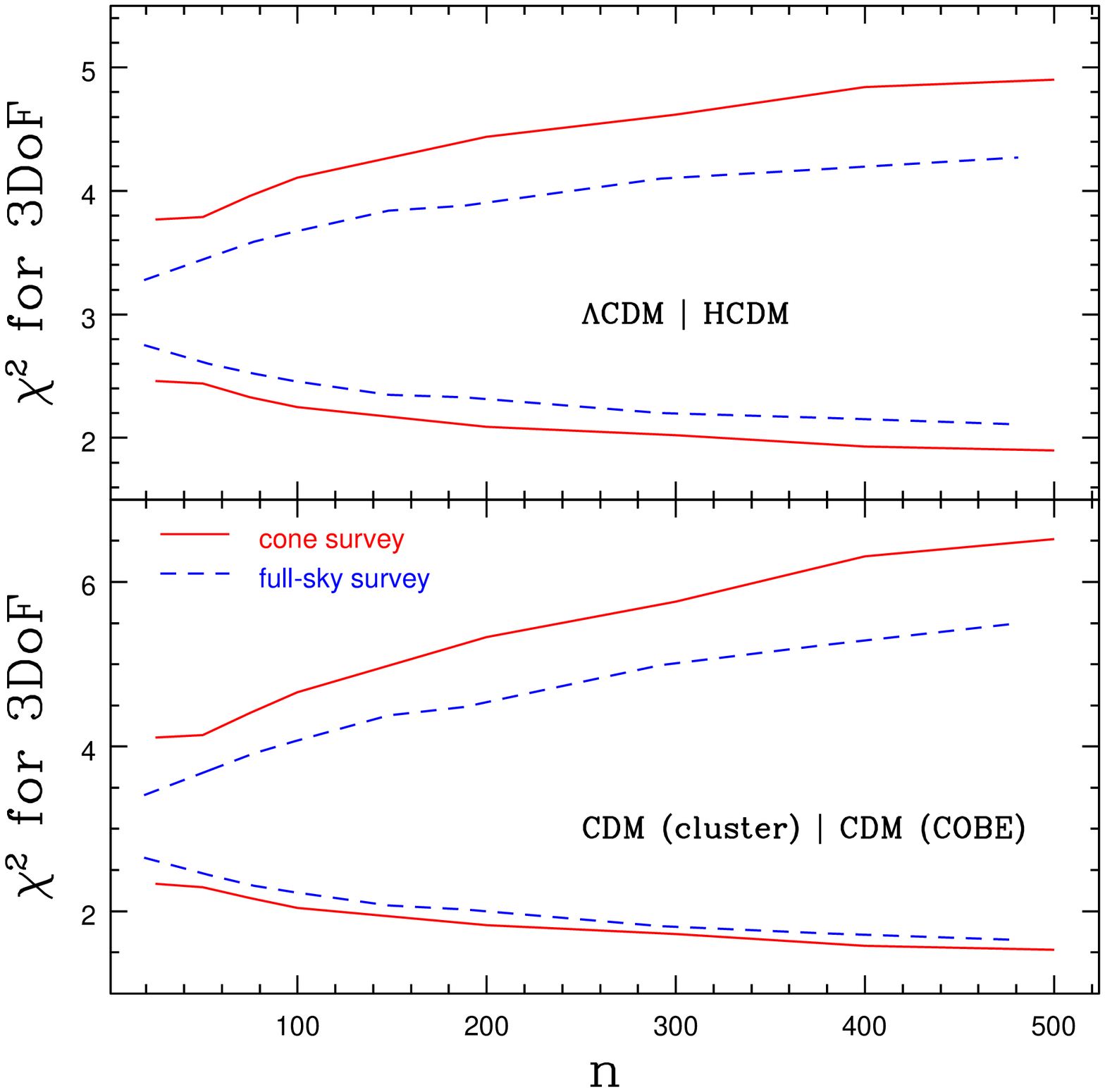} \caption{\figurethree}
	\label{fig3} \end{figure}
\fi

In the upper panel of Figure 3 we show the $\chi^2$ for three degrees of
freedom for distinguishing between HCDM and
$\Lambda$CDM in a $10^{\circ}$ cone survey with $R_o=100$\hmpc\ as a
function of the number of objects.  The upper curve is for HCDM assuming
that the true power spectrum is $\Lambda$CDM; here the $\chi^2 > 3$
since we are testing a model with excess power relative to the actual
spectrum.  The lower curve is for $\Lambda$CDM assuming the true
spectrum is HCDM, which will give $\chi^2 < 3$.  From these $\chi^2$
values, one can assign a confidence level at which the models are ruled
out.  In general, one can rule out spectra
with excess power at a higher confidence level.  For comparison, we show
the same quantities for a full-sky survey (with the zone-of-avoidance
removed) using the three bulk flow components as the three degrees of
freedom; these surveys have the same radial distribution, number of
objects and errors as the cone surveys.  In the lower panel we show the
$\chi^2$ for COBE normalized CDM given cluster normalized CDM and vice
versa.  From the figure we see that the three moments calculated for the
cone survey do better than the three bulk-flow components of the
full-sky survey at distinguishing between the models for the same number
of observed galaxies.

A more complete comparison between full-sky and cone surveys
would include information from higher moments of the full-sky
survey.  However, for the surveys and power spectra we have
considered, the signal to noise of these higher moments
is small. Results from analyses including
the highest signal to noise moments suggest that cone surveys
continue to be as good or better at distinguishing spectra.  

\section{Conclusions}

In this {\it Letter} we have explored the properties of proper distance
surveys that cover small fields in two opposing directions.  Our
analysis exploits the fact that a small area survey can measure some of
the moments of the velocity field much more accurately than a full sky
survey with the same number of objects.  We have shown how to expand the
velocity field in moments and constructed a $\chi^2$ test useful for
constraining models. In order
to get a ``signal to noise'' of unity for the three lowest moments, we
found that a survey of $\approx 200-300$ galaxies is needed if we assume
distance indicators accurate to about $15\%$ of the distance in a cone
survey with opening angle of ${\cal O}(10^\circ)$ and consider survey
depths $\sim200$\hmpc.  We have also shown that cone surveys are as good
or better than full-sky surveys at distinguishing between cosmological
models.  These surveys could put
important constraints on the power spectrum on large scales with only a
modest observational effort, and thus could provide a valuable tool in
probing scales that have been up to now largely beyond our scope.

\noindent{\bf Acknowlegements:} We would like to thank Nick Kaiser for
many conversations. We would also like to thank Michael Strauss, Gary
Wegner and Jeff Willick for many
thoughtful comments. HAF was supported in part by the NSF EPSCoR Grant
and the University of Kansas GRF.  RW was supported in part by NSF grant
PHY-9453431 and NASA grant NAGW-4720.

\vfill\eject

\ifnum\putfigure>0
	\begin{figure} \caption{\figureone \label{fig1}} \end{figure}
	\begin{figure} \caption{\figuretwo \label{fig2}} \end{figure}
	\begin{figure} \caption{\figurethree \label{fig3}} \end{figure}
\fi

\vfill\eject

\end{document}